\documentclass[prb,aps,twocolumn,showpacs,home,floats]{revtex4}
\usepackage{graphics}
\begin{document}

\title{
{\rm\small\hfill (in press, Phys. Rev. Lett.)}\\
Why is a noble metal catalytically active?
The role of the O-Ag interaction in the function of
silver as an oxidation catalyst}

\author{Wei-Xue Li$^{1}$, Catherine Stampfl$^{1,2}$, and 
Matthias Scheffler$^{1}$}
\affiliation{$^{1}$Fritz-Haber-Institut der Max-Planck-Gesellschaft, Faradayweg
4-6, D-14195 Berlin-Dahlem, Germany\\
$^{2}$Department of Physics and Astronomy, Northwestern University,
Evanston, Il. 60208}
\date{Received 12 August 2002}  

\begin{abstract}
Extensive density-functional theory calculations, and
taking into account temperature and pressure,
affords a comprehensive picture of the behavior and interaction
of oxygen and Ag(111), and provides valuable insight into the
function of silver as an oxidation catalyst.
The obtained phase-diagram reveals the most stable species present
in a given environment and thus identifies (and excludes)
possibly active oxygen species.
In particular, for the conditions of
ethylene epoxidation, a thin oxide-like structure is most stable,
suggesting that such atomic O species are actuating  the catalysis, 
in contrast to 
hitherto proposed molecular-like species.
{\em Copyright (2003) by The Americal Physical Society.}
\end{abstract}

\pacs{PACS numbers: 81.65.Mq, 68.43.Bc, 82.65.My}

\maketitle

The importance of reactant bond strength 
in heterogeneous catalysis
is demonstrated by the so-called ``volcano curves''~\cite{cataly-2,Balandin69,sab}
which relate the reaction rate 
to the adsorption energy of the reaction intermediates. 
Typically a maximum occurs in the rate at moderate values.
This behavior underlies the expectation that
a good catalyst should readily dissociate adparticles
but not bind the fragments too strongly -- a concept put forward by
Sabatier in the early 1900's, which 
recent theoretical and experimental studies
have re-addressed.~\cite{norskov,toulhoat}
Despite this expectation, silver, as a noble metal, 
which binds adparticles only relatively weakly on the surface,
is a very important heterogeneous catalyst for various
oxidation reactions.
For example, partial oxidation of methanol to formaldehyde,
as carried out at atmospheric pressure and  temperatures 
of 800-900~K,\cite{nagy,here96} as well as
the selective oxidation of ethylene to epoxide,
also conducted 
at atmospheric pressure and   
at temperatures of 500-600~K.~\cite{sant87}

Although numerous
investigations aimed at elucidating the active oxygen species
involved in the above-mentioned reactions have been
carried out, it can be said that presently there
is considerable confusion, and no unambiguous
identifications have been made.
For example, unclear issues concern
the debate of molecular (ozone-like)~\cite{boron}
versus atomic~\cite{bukh011} oxygen as the active species
for ethylene epoxidation, the proposed importance 
of ``bulk-dissolved'' oxygen and a ``strongly bound, surface-embedded''
species (which desorbs at 900~K) for the
partial oxidation of methanol,~\cite{nagy} as well as
the possible role
of chemisorbed or surface oxide-like species.
Part of the reason for this lack of understanding 
is due to the  difficulties associated with the pressure and materials
gaps which exist
for the oxygen--silver system. That is, the behavior 
is different under ultra-high vacuum (UHV) conditions, where the
system can be
analysed quantitatively, to that under the high temperature and pressure
conditions of catalysis, 
where it is much more difficult to obtain the same level
of microscopic information.

Our theoretical strategy, aimed at gaining an understanding of the
function of silver as an oxidation catalyst, in spite
of its relatively weak binding capability, 
and to identify (and
exclude) possibly active species, is to carry out extensive
and systematic
density functional theory (DFT) calculations for 
all conceivably relevant oxygen
species.
We take the effect of temperature ($T$) and pressure ($p$)
into account through
the oxygen chemical potential, and thus determine the 
($T,p$) phase-diagram which describes the surface phases
from UHV right up to the conditions of real catalysis.
We study oxygen at the (111) surface of silver as experiments
indicate that this orientation is an important crystal face
for real silver catalysts in that at high temperatures,
facets   with this face result.~\cite{nagy,bao} 
Our general findings, however,
are expected to be relevant for silver {\em per se}, and possibly
to have implications for gold, 
also a noble metal oxidation catalyst~\cite{gold}
which, for the (111) surface, exhibits a restructuring
and chemisorption of oxygen atoms at elevated temperatures
(500-800~K) and 
atmospheric pressures,~\cite{gold1,gold2}
similar to Ag(111).
They could also have consequences for copper, 
which catalyses the oxidation of methanol to formaldehyde.~\cite{copper}

Our calculations employ 
the pseudopotential~\cite{trou91,fuchs99}
plane-wave method~\cite{fhi98} with the generalized
gradient approximation (GGA)~\cite{pbe96} for the exchange-correlation
functional. Five Ag layers are used in the supercell to model the 
O/Ag surfaces,
which are created on one side of the slab,
with a vacuum region of 15~\AA\,. 
The same {\bf k}-point sampling
is  used for all structures. It corresponds to 21 points in 
the irreducible part of the 
Brillouin zone of the $(1\times1)$ Ag(111) surface cell.
An energy cutoff of 50~Ry is used and
full relaxation of the top two or three silver 
layers, and the O atoms, is taken into account.
We include the spin-polarization energy in calculation
of the total energy of the free atoms and molecules.
Further details 
are reported in Ref.~\onlinecite{wxli01}.

The oxygen species investigated, for a wide range of coverages, include:
on-surface sites, surface- and bulk-substitutional sites,
interstitial sites under the surface Ag layer and deeper in the
bulk, as well as oxide-like structures, 
and a molecular ozone-like species adsorbed at a surface Ag vacancy.

\begin{figure}
\scalebox{0.70}{\includegraphics{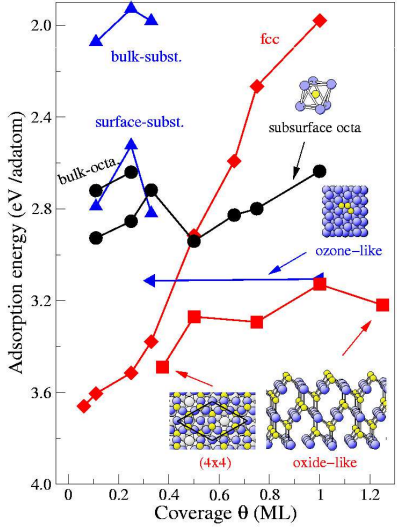}}
\caption{Selected results of average adsorption energies (with respect
to atomic O) versus coverage for various 
structures:
pure on-surface oxygen in fcc-hollow sites (red diamonds),
surface- and bulk-substitutional adsorption (blue  triangles),
pure sub-surface oxygen in octahedral
sites under the first (``subsurface-octa.'')
and second (``bulk-octa.'') Ag layers (black circles), an O$_{3}$-like molecule
adsorbed at a surface Ag vacancy
(left-pointing black triangles), and 
oxide-like structures (red squares). 
Oxygen atoms are depicted as yellow spheres
and Ag atoms as the larger blue (and grey)
spheres.
}
\label{fig:ave-energies}
\end{figure}

The main important adsorption 
energies 
are summarized in Fig.~1, where for
each ``type'' of oxygen atom, only the lowest energy structure is shown.
For on-surface adsorption, 
the fcc-hollow site is preferred
for all coverages investigated. 
For sub-surface adsorption, 
the octa. site under the first Ag layer is the most favorable.
This is also the case for adsorption under the second Ag layer
which may be taken to represent ``bulk dissolved'' oxygen.
For all sub-surface  sites,
adsorption under the second layer is {\em less} favorable
than under the first layer.
In view of the unfavorable energy of oxygen in the bulk,
high concentrations of bulk-dissolved oxygen in the
perfect fcc lattice of silver are  improbable.

Surface- or bulk-substitutional adsorption is
seen from Fig.~1 to be unfavorable, as is the
ozone-like species.
To investigate the formation of oxide-like structures,
we performed systematic investigations for
geometries involving  coverages of 0.50, 0.75, 1.0, and 1.25~ML using
$(2\times 2)$ cells.
In particular, we tested 
combinations of on-surface fcc and hcp sites, and the three
possible sub-surface sites under the first Ag layer.
The atomic geometry of the favored structure
consists of 0.25~ML of oxygen on the surface and 0.5~ML 
in between the Ag layers. 
For increasing O concentrations, this structure ``grows'' by filling deeper
lying sites between the Ag layers.
The structure for a total coverage of 1.25~ML oxygen  is
sketched in the inset of Fig.~1.
In addition, we carried out
calculations for the recently proposed $(4\times 4)$ thin ``surface-oxide'' 
structure~\cite{king}
which involves an O-Ag-O bonding unit (and electronic
structure) similar to the upper
tri-layer of the oxide structures described
above (see insets of Fig.~1).
The average adsorption
energies of all these oxide-like structures are displayed in Fig.~1, shown as red squares.
Their atomic geometries are in fact
very similar to the (111) surface of Ag$_2$O.

\begin{figure}
\scalebox{0.43}{\includegraphics{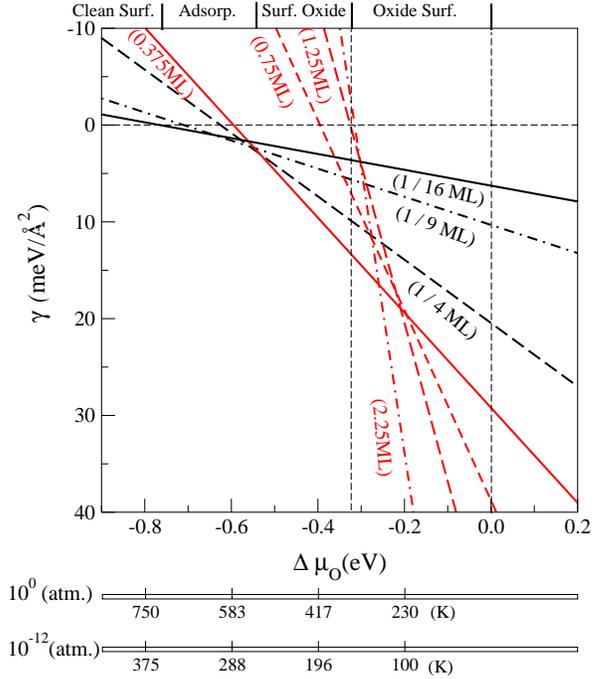}}
\caption{Surface free energies 
for various low energy structures as a function
of the O chemical potential where $\Delta \mu_{\rm O}=
\mu_{\rm O}-\frac{1}{2}E_{\rm O_2}^{\rm total}$. 
For pressures of 10$^{-12}$ atm. (UHV)
and 1 atm. (condition of catalysis), 
the corresponding temperatures are given. 
The temperature labels correspond to the tick marks
and labels
for the chemical potential.
The labels ``0.75~ML'', ``1.25~ML'',
and ``2.25~ML'', indicate the O-concentrations in the 
corresponding oxide-like structures. 
Accordingly the $(4\times4)$ structure has
the label 0.375~ML, and the adsorption structures 1/4, 1/9, and 1/16~ML.
At the top of the figure, the ``material type'' which is
stable in the corresponding range of chemical potential are listed
and indicated by the colored regions.
}
\end{figure}
We turn now to investigate the effect of 
pressure and temperature on the stability of the
various structures.
To do this we calculate the surface free energy, 
\begin{eqnarray}
\gamma(T,p)=(G - N_{\rm Ag} \mu_{\rm Ag}
- N_{\rm O}  \mu_{\rm O}) / A \quad ,
\label{eq-gibbs}
\end{eqnarray}
\noindent
where $G=G^{\rm slab}_{\rm O/Ag(111)} - G^{\rm slab}_{\rm Ag(111)}$
and the first and second terms on the right-hand-side 
are the free energies of the O/Ag surface under
consideration and that of the reference system,
i.e. the clean Ag(111) slab, respectively.
$A$ is surface area in \AA\,$^{2}$ and
$N_{\rm O}$ is the number of oxygen atoms. $N_{\rm Ag}$ is
the number of silver atoms (with respect to
the reference system) and $\mu_{\rm Ag}$ is
the Ag chemical potential, which is the free energy
of an Ag atom in bulk fcc silver. (Consideration of
$N_{\rm Ag}$ and $\mu_{\rm Ag}$ are necessary only for the
substitutional structures and the $(4\times4)$ phase).
The $T$ and $p$ dependence is given by $\mu_{\rm O}$, the
oxygen chemical potential,\cite{reuter} 
\begin{eqnarray}
\mu_{\rm O}(T,p) = 1/2[E_{\rm O_{2}}^{\rm total} +
\tilde{\mu}_{\rm O_2}(T,p^{0})
 \;+\;  \;k_{B} T
  \;{\rm ln} \left( \frac{p_{\rm O_2}}{p^{0}} \right)],
  \label{eq:ochem}
\end{eqnarray}
where $p^{0}$ corresponds to atmospheric pressure and
$\tilde{\mu}_{\rm O_2}(T,p^{0})$ includes the contribution from
rotations and vibrations of the molecule, as well as the
ideal-gas entropy at 1 atmosphere. Here we use
experimental values from thermodynamic tables.

Through consideration of the difference in the 
vibrational and entropic  contributions of $G^{\rm slab}_{\rm O/Ag(111)}$
and $G^{\rm slab}_{\rm Ag(111)}$, we find that it  
can be neglected.~\cite{wxli03}
Thus, for evaluating {\em the
difference} of the two $G^{\rm slab}$ values, we replace
$G^{\rm slab}$ by the total energy.
We choose the zero of  the O chemical potential
to be half the total energy of O$_{2}$ calculated as
 $1/2E_{\rm O_2}^{\rm total} \simeq E_{\rm Ag_2O}^{\rm total} \; - \;2
E_{\rm Ag}^{\rm total}
\; - \; H^{f}_{\rm Ag_2O-bulk}$. That is, as the energy difference
between bulk silver oxide ($E_{\rm Ag_2O}^{\rm total}$) and bulk silver  
(2$E_{\rm Ag}^{\rm total})$,
and the experimental value of the heat of formation 
($H^{f}_{\rm Ag_2O-bulk}$) which is $-0.323$~eV.~\cite{crc}

The obtained results are displayed in Fig.~2.
The range of $\Delta \mu_{\rm O}$ (defined as
$\mu_{\rm O}-\frac{1}{2}E_{\rm O_2}^{\rm total}$)
between the vertical dashed lines 
corresponds to the heat of formation of 
Ag$_{2}$O, i.e., to the range in which
bulk silver oxide is stable.
It can be seen that for more O-rich conditions
(right) the thicker oxide-like structures are favored over chemisorption
on the surface.  
Clearly, in the limit of an infinitely thick film, the corresponding
line would be very close to the vertical one 
at $\Delta \mu_{\rm O}=-0.323$~eV.
It can be obtained 
from Fig.~2 that bulk silver oxide is stable to only around 350~K at
atmospheric pressure, 
which is less than the experimental value of 460~K. 
At 460~K, the bulk oxide is unstable when the pressure is lower
than 10$^{3}$ atm. i.e. compared to 1 atm. as observed experimentally.
This
difference may be due to {\em systematic} errors of the DFT 
approach~\cite{comment}
and/or to neglect of the entropy contributions.
Therefore in Fig.~2 (and Fig.~3), we generally expect that the
temperature is underestimated, but the error is less than 110~K,
while correspondingly, the pressure is overestimated.
For the temperature values quoted below, we give an estimate
of the upper limit in brackets,
that is, 110~K higher than that obtained
from Fig.~2.
In spite of these uncertainties, we believe that
the relative stability of the various systems and our
general understanding and conclusions will not be affected.

Due to the above-mentioned low thermal stability of silver oxide and the 
thicker oxide-like structures,
our results can safely rule them out
as playing an important role in the catalytic reactions.
For values of the O chemical potential further to the left in Fig.~2,
and atmospheric pressure,
the $(4\times4)$ phase is the most stable
for the temperature range 350-530~K (incl. corrections, 460-640~K).
For higher temperatures, up
to about 720~K (incl. corrections, 830~K), 
on-surface adsorbed oxygen is the only stable
species. For temperatures beyond this, there are no stable species 
except for O atoms adsorbed at under-coordinated surface Ag atoms,
such as next to a vacancy or step edge, which are
stable to 865~K (incl. corrections, 975~K).
From the above analysis, our results point to 
oxygen atoms of the thin $(4\times4)$ surface-oxide structure
as the main species actuating ethylene epoxidation as this occurs
experimentally at $p_{\rm O_{2}}=1$~atm. and $T=500$-600~K, while
either low coverages of on-surface oxygen or oxygen adsorbed
at defects may play a role in the higher temperature catalytic
reactions.

In Fig.~3, we consider just the lowest energy structures
as a function of temperature
($T$) and pressure ($p$), yeilding the phase-diagram. 
Here the $p$-$T$ range in which the various
structures are stable can quickly be seen.
We note that the phase-diagram is in-line with 
available well-established experimental
results, namely, (i) that under UHV conditions, only very low coverage
(0.03~ML)
of chemisorbed O is stable up to about 490~K and (ii)
that the $(4\times4)$ phase is 
stable at low temperatures (and UHV pressure) and that it decomposes
at 580~K.~\cite{wxli03}
Figure 3 furthermore shows that
regardless of the pressure, the ordering of the phases 
is the same. Thus, the same phase should be observable at different pressures
if the temperature is adjusted accordingly, which indicates a way to
bridge the pressure gap -- providing that formation of the structure is
not prevented by kinetics.

It is interesting to consider the ``removal'' energy
of a single (uppermost) O atom
of the $(4\times4)$ structure,
which is the energy required to move the O atom into the vacuum. 
This energy is the strongest that we found for the O/Ag system, namely,
$\sim$3.9~eV, which is due to the presence of the lower lying O atoms
which stabilizes the upper O atoms.
Compared to other O/metal systems, however, this does not constitute a strong
bond. In fact, it represents an {\em intermediate}
value, very  similar to the identified ``optimum'' 
values for highest catalytic
activity of certain reactions.~\cite{norskov}
Interestingly, recent studies of the CO oxidation reaction over Ru
have surprisingly found that oxides play an active role      
for reaction, in contrast to the hitherto believed pure 
metal.~\cite{over,ss-500}
Ruthenium is a poor oxidation catalyst since it binds
O too strongly, but at the high temperature and pressure conditions 
where Ru-oxides form, 
a new weaker O species appears and the system is then catalytically 
very active.
Thus it appears that the interaction of oxygen with transition metal
catalysts can serve to ``tune'' the O-metal bondstrength 
for optimum reactivity through formation of various oxide-like species.
\begin{figure}
\scalebox{0.38}{\includegraphics{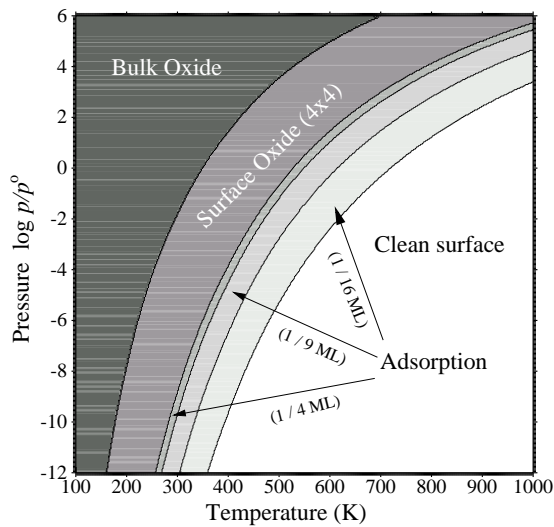}}
\caption{Calculated $(p,T)$ phase-diagram for 
the oxygen-Ag(111) system showing the stable structures.}
\label{phase-bulk-new}
\end{figure}

In conclusion, through DFT calculations,
and including
the effect of the environment, we obtained the pressure--temperature 
phase-diagram for O/Ag(111).
The results show that 
a proposed molecular ozone-like species adsorbed at a surface vacancy is
energetically {\em unfavorable}, as is bulk-dissolved oxygen, which
puts a question mark over the latter's hitherto thought important role
in oxidation reactions.
Our results also show that
silver oxide and oxide-like films can also be ruled out as playing
an important role due to their low thermal stability.
On the other hand, the calculations reveal that
silver can support
a {\em thin} surface oxide-like structure under
the temperature and pressure conditions of ethylene epoxidation and
we propose that such atomic O species are actuating  the reaction.
For even higher temperatures,
only very low concentrations of chemisorbed O are stable, or
O atoms adsorbed at under-coordinated surface Ag atoms.
Therefore these species could possibly play a role in the
catalytic oxidation reactions that take place at high temperature.
We note that an understanding of a {\em full catalytic cycle}
requires a kinetic modeling that includes all reactant species
and intermediates, which, for complex systems such
as those discussed here, is not yet possible.
The present investigation represents a crucial first
step towards this aim, which, using the approach
of atomistic ab initio thermodynamics, identifies (and
excludes) oxygen species that
are present under the conditions of catalysis, and helps shed
new light upon the function of silver as
an oxidation catalyst.

More generally, our results highlight the formation and
importance of various oxides under catalytic conditions, be it 
in weaker bonded O/noble-metal systems, such as O/Ag, or in 
stronger bonded O/transition-metal systems, as e.g.,
O/Ru.  For transition metals that form bulk oxides
with intermediate
heats of formation, e.g. PdO, under conditions typical of catalysis,
we may expect that both bulk oxides and thiner surface-oxides could occur,
depending sensitively on the pressure and temperature.
\\

We thank D. King and R. Schl\"{o}gl for helpful discussions.

\end{document}